\documentclass[twocolumn,nofootinbib,showpacs,superscriptaddress,fleqn]{iopart}

\usepackage{amsfonts}
\usepackage{amstext}
\usepackage{amssymb}
\usepackage{bm}
\usepackage{cite}
\usepackage{dcolumn}
\usepackage{graphicx}
\usepackage{graphics}
\usepackage[latin1]{inputenc}
\usepackage{latexsym}
\usepackage{rotating}
\usepackage{url}
\usepackage[colorlinks=true]{hyperref}
\usepackage{xspace} 
\usepackage[usenames]{color}
\usepackage{booktabs}

\definecolor {darkgreen}{rgb}{0.2,0.7,0.2}

\newcommand\be{\begin{equation}}
\newcommand\ba{\begin{eqnarray}}
\newcommand\ee{\end{equation}}
\newcommand\ea{\end{eqnarray}}

\newcommand\bw{\begin{widetext}}
\newcommand\ew{\end{widetext}}

\newcommand{\K}{{\mbox{\tiny K}}}

\newcommand{\QG}{{\mbox{\tiny QG}}}
\newcommand{\EdGB}{{\mbox{\tiny EdGB}}}
\newcommand{\dCS}{{\mbox{\tiny dCS}}}

\newcommand{\MAT}{{\mbox{\tiny mat}}}

\newcommand{\ISCO}{{\mbox{\tiny ISCO}}}

\begin{document}
\title[Black Hole Continuum Spectra as a Test of General Relativity]{Black Hole Continuum Spectra as a Test of General Relativity: Quadratic Gravity}

\author{%
Dimitry~Ayzenberg$^{1}$
and
Nicol\'as~Yunes$^{1}$
}

\address{$^{1}$~eXtreme Gravity Institute, Department of Physics, Montana State University, Bozeman, MT 59717, USA.}

\date{\today}

\begin{abstract} 

Observations of the continuum spectrum emitted by accretion disks around black holes allows us to infer their properties, including possibly whether black holes are described by the Kerr metric. 
Some modified gravity theories do not admit the Kerr metric as a solution, and thus, continuum spectrum observations could be used to constrain these theories. 
We here investigate whether current and next generation X-Ray observations of the black hole continuum spectrum can constrain such deviations from Einstein's theory, focusing on two well-motivated modified quadratic gravity theories: dynamical Chern-Simons gravity and Einstein-dilaton-Gauss-Bonnet gravity. 
We do so by determining whether the non-Kerr deviations in the continuum spectrum introduced by these theories are larger than the observational error intrinsic to the observations. 
We find that dynamical Chern-Simons gravity cannot be constrained better than current bounds with current or next generation continuum spectrum observations. 
Einstein-dilaton-Gauss-Bonnet gravity, however, may be constrained better than current bounds with next generation telescopes, as long as the systematic error inherent in the accretion disk modeling is decreased below the predicted observational error.

\end{abstract}

\submitto{\CQG}

\maketitle

\section{Introduction}

The recent detections of gravitational waves from binary black hole (BH) mergers by advanced LIGO (aLIGO)~\cite{PhysRevLett.116.061102, Abbott:2016nmj} have ushered in the era of \emph{extreme gravity} tests of General Relativity (GR)~\cite{TheLIGOScientific:2016src,Yunes:2016jcc}, i.e.~probes that sample the non-linear and dynamical nature of the gravitational interaction. In the coming years, these two observations will be bolstered by further gravitational wave detections from aLIGO, aVirgo, and KAGRA~\cite{PhysRevD.88.043007} and pulsar timing arrays (PTAs)~\cite{lrr-2003-5}. In the meantime, electromagnetic (EM) observations of accretion disks around BHs using new X-Ray telescopes and very long baseline interferometers (VLBIs)~\cite{Takahashi:2014cea, Zhang:2016ach, Fish:2009va, Goddi:2016jrs, lrr-2008-9, 2015arXiv150903884B} will also join the game. Unlike gravitational wave observations, EM ones do not directly probe the dynamical sector of the gravitational interaction, but instead, they detect the impact of a stationary source of strong gravity on radiation that attempts to escape from it. These observations include, but are not limited to, the continuum spectrum of BH accretion disks, the BH ``shadow'', and the K$\alpha$ iron line emitted from BH accretion disks~\cite{Fish:2009va, 1538-4357-482-2-L155, 2000ApJ...528L..13F, Bambi:2016sac}.

Continuum spectrum observations are of particular interest as the BH accretion disk spectrum is very sensitive to the properties of the BH spacetime. The continuum spectrum of thin disks is dominated by radiation originating near the inner radius of the accretion disk, which can be well approximated by the innermost stable circular orbit (ISCO)~\cite{2015arXiv150903884B}, i.e.~the last stable orbit an accretion disk (test) particle can be in before plunging into the BH. The ISCO is independent of the accretion disk properties and depends only on the properties of the BH spacetime, e.g.~the BH mass $M$ and (the magnitude of) the spin angular momentum $|\vec{J}|$. Continuum spectrum observations are thus a useful tool for determining the properties of BHs and have been used to estimate the angular momentum of several BHs~\cite{2015PhR...548....1M}.

Black hole continuum spectrum observations could allow us, at least in principle, to test the \textit{Kerr hypothesis}, i.e.~that all isolated, stationary, and axisymmetric astrophysical (uncharged) BHs are described by the Kerr metric~\cite{1975PhRvL..34..905R, 1967PhRv..164.1776I, 1968CMaPh...8..245I, 1971PhRvL..26.1344H, 1972CMaPh..25..152H, 1971PhRvL..26..331C}, which is completely determined by only two parameters (the mass and the angular momentum). The Kerr metric is a solution to the Einstein equations in vacuum, but it can also be a solution in certain modified gravity theories~\cite{Psaltis:2007cw}. In this sense, verifying the Kerr hypothesis is a null test of General Relativity, but it does not necessarily rule out all modified gravity models. There are, however, modified theories of gravity that introduce violations to fundamental pillars of GR, and that, in particular, do not satisfy the Kerr hypothesis. In these cases, BH continuum spectrum observations can be used to place constraints on these theories.

Two such modified gravity theories that have also been well-studied are dynamical Chern-Simons (dCS) gravity~\cite{Alexander:2009tp} and Einstein-dilaton-Gauss-Bonnet (EdGB) gravity~\cite{Moura:2006pz}. Both theories modify the Einstein-Hilbert action by introducing a dynamical scalar field that couples to a curvature invariant, the Pontryagin invariant in the case of dCS gravity and the Gauss-Bonnet invariant in the case of EdGB gravity. These theories break parity invariance in the gravitational sector and the strong equivalence principle, both of which are pillars of Einstein's theory~\cite{lrr-2014-4}. Black holes within these theories have been studied extensively and many numerical and approximate solutions have been found, although no exact solution is known for BHs that spin arbitrarily fast~\cite{Campbell:1990fu, Campbell:1990ai, Campbell:1991kz, Campbell:1992hc, Alexeev:1996vs, Kanti:1995vq, Torii:1996yi, Kleihaus:2011tg, PhysRevD.83.104002, Yunes:2009hc, Pani:2011gy, McNees:2015srl}. For the purpose of this work we will focus on a pair of purely analytic, approximate solutions, one for each theory. Both solutions are stationary and axisymmetric, and they represent a slowly-rotating BH to quadratic order in the ratio of spin angular momentum to BH mass squared~\cite{Ayzenberg:2014aka, PhysRevD.86.044037}. 

An overarching goal of our research program is to determine the degree to which BH continuum spectrum observations can be used to constrain deviations from GR. A step in this direction and the primary goal of this paper is to consider constraints on dCS gravity and EdGB gravity. Both theories have been constrained with observations that are not in the extreme gravity regime~\cite{Kent-LMXB, alihaimoud, PhysRevD.86.044037}, and thus, they are not very stringent. Given that EM observations are sensitive to strong-field physics, one could expect that they may lead to much more stringent constraints, if the modified gravity effects are not overwhelmed by observational error and modeling systematics. In this paper we will investigate this topic with continuum spectrum observations using both current and next generation X-Ray telescopes. 

Whether continuum spectrum observations can be used to place constraints on modified gravity will depend on whether the induced deviations in the spectrum are \emph{detectable}. As a proxy for detectability, we will here require that at the very least the modified gravity deviations should be larger than any other systematic, statistical, instrumental, and environmental error in the observations. Systematic error originates from the approximate nature of the models we use to analyze the data, in particular the approximate nature of BH solutions in modified gravity and of the astrophysical models for accretion disks. The impact of using approximate, slowly-rotating BH solutions for continuum spectrum observations was studied recently~\cite{Ayzenberg:2016ynm}, with results suggesting that the systematic error introduced is negligible, provided the BH is not close to maximally rotating. The impact of using approximate accretion disk models is currently unknown, because of their complexity and the large number of proposed models. In fact, there is much debate over which model(s) best describes continuum spectrum observations in GR~\cite{lrr-2013-1, Lasota:2015cga}. This source of systematic error is beyond the scope of this paper. Statistical, instrumental, and environmental errors will be collectively referred to as \textit{observational errors}, and they affect the accuracy to which BH properties can be estimated by roughly $10\%$ with current X-Ray telescopes~\cite{2015PhR...548....1M}; next generation X-Ray telescopes will be able to estimate BH properties to $1\%$~\cite{Takahashi:2014cea, Miller:2014cda, Zhang:2016ach}

In order to determine whether continuum spectrum observations can be used to place constraints on modified gravity, we perform a parameter estimation study in which we treat the continuum spectrum of a Kerr BH as the \textit{observation} or \textit{injection} and the continuum spectrum of a BH in EdGB gravity or dCS gravity as the \textit{model} we fit to the injection. The parameters we estimate are the BH mass, the BH spin angular momentum, and the inclination angle, i.e.~the angle between the observer's line of sight and the BH's angular momentum. The parameter estimation study is done by minimizing the relative $\chi^2$ over all parameters. We model the observational error using errors in the parameters that are representative of current and next generation continuum spectrum observations. Since the model is different from the observation, our parameter estimation will result in biased best-fit parameters that will differ from the true parameters of the signal. Continuum spectrum observations can then be used to constrain EdGB and dCS gravity provided the difference between the biased parameter and the true parameters are larger than the observational errors.

The main results of this paper are that dCS gravity and EdGB gravity cannot be constrained better than current constraints using continuum spectrum observations with current X-Ray telescopes. Assuming a signal consistent with GR, the bias in the recovered BH parameters when analyzing the data with dCS gravity and EdGB gravity models is much smaller than the observational error with current telescopes. Using next generation telescopes, on the other hand, it may be possible to place better than current constraints on EdGB gravity, but not on dCS gravity. With a reduction in the observational error of an order of magnitude, the bias in parameter extraction with EdGB gravity models becomes significantly larger than the observational error for BHs of mass $M\lesssim 8M_\odot$. This is not, however, the case with dCS gravity models, for which the bias in parameter extraction remains well below the observational error. This is, in part, because non-spinning BHs are still described by the Schwarzschild metric in  dCS gravity, with modifications only important very close to the BH's event horizon and always proportional to the spin. These results are summarized in Table~\ref{table:summary}.

\begin{table*}[hpt]
\centering
\begin{tabular}{l c c}
\toprule
& Current & Next Gen.~
\\ \midrule
dCS & $\times$ & $\times$
\\
EdGB & $\times$ & $\checkmark$
\\ \bottomrule
\end{tabular}
\caption{Table summarizing main results of this paper, i.e.~whether observations of black hole continuum spectra with current and next generation X-Ray telescopes can be used to place better than current constraints on dynamical Chern-Simons gravity and Einstein-dilaton-Gauss-Bonnet gravity. A checkmark, $\checkmark$, means better than current constraints can be placed $\forall~M_{\rm\tiny{ BH}}\lesssim8M_\odot$, while a cross, $\times$, means better than current constraints cannot be placed. \label{table:summary}}
\end{table*}
%

As part of this investigation, we also obtain a set of secondary results. We calculate properties of EdGB and dCS gravity BH solutions that play an important role in the calculation of the spectrum, such as the conserved energy and angular momentum, the ISCO radius, and the gravitational redshift. More importantly, perhaps, we also show that test particles, i.e.~particles with extremely weak self-gravity, follow geodesics in EdGB gravity, developing a proof that is  similar to that employed previously in dCS gravity~\cite{Sopuerta:2009iy}. This proof allows one to continue to use the geodesic equation to solve for the motion of accretion disk particles in the background of a massive BH.

EdGB gravity and dCS gravity have been studied extensively in the context of testing GR~\cite{Kent-LMXB, alihaimoud, PhysRevD.86.044037, 2015CQGra..32u4002S, Cardoso:2015xtj, Yagi:2016jml, Chakraborty:2012sd, Pani:2011xm, Sotiriou:2014pfa}. Our work extends previous work on BH electromagnetic observations in EdGB gravity and dCS gravity by studying the continuum spectrum of solutions in these theories that have not been studied in the past. Although a similar study had been carried out in dCS gravity before~\cite{Vincent:2013uea}, this was only to linear-order in spin. DCS modifications that are quadratic in the spin enter the diagonal components of the metric, which could in principle have a larger effect on electromagnetic observables. The continuum spectrum in EdGB gravity had not been studied before, although other electromagnetic observables had been considered, such as the black hole shadow~\cite{2017arXiv170100079C} and quasi-periodic oscillations~\cite{Maselli:2014fca}. Similar results to those in this work were found in the case of quasi-periodic oscillations, namely that next generation telescopes should be able to place better than current constraints on EdGB gravity. Showing that this result holds for multiple types of electromagnetic observations adds to the scientific case of next generation telescopes.

The remainder of this paper presents the details of the calculations that led us to the above conclusions. Section~\ref{sec:QG} briefly summarizes the quadratic gravity (QG) class of modified gravity theories of which dCS gravity and EdGB gravity are a part and the BH solutions this paper studies. Section~\ref{sec:prop-of-sol} presents properties of the BH solutions that are relevant to the continuum spectrum calculation. Section~\ref{sec:spec} details the continuum spectrum calculation, our statistical analysis methodology, and the results obtained. Section~\ref{sec:disc} concludes by summarizing our results and discussing implications. Throughout, we use the following conventions: the metric signature $(-,+,+,+)$; Latin letters in index lists stand for spacetime indices; parentheses and brackets in index lists for symmetrization and antisymmetrization, respectively, i.e. $A_{(ab)}=(A_{ab}-A_{ba})/2$ and $A_{[ab]}=(A_{ab}-A_{ba})/2$; geometric units with $G=c=1$ (e.g. $1 M_\odot$ becomes 1.477 km by multiplying by $G/c^2$ or $4.93\times10^{-6}$ s by multiplying by $G/c^3$), except where otherwise noted.

\section{Quadratic Gravity and BH Solutions}
\label{sec:QG}

The action that describes the QG class of theories is defined by a modification to the Einstein-Hilbert action containing all possible quadratic, algebraic curvature scalars with running (i.e.~nonconstant) couplings~\cite{PhysRevD.83.104002}
\begin{eqnarray}
S&\equiv\int d^4x\sqrt{-g}\{\kappa R+\alpha_1f_1(\vartheta)R^2+\alpha_2f_2(\vartheta)R_{ab}R^{ab}
\nonumber \\
&+\alpha_3f_3(\vartheta)R_{abcd}R^{abcd}+\alpha_4f_4(\vartheta)R_{abcd}\text{}^{*}R^{abcd}
\nonumber \\
&-\frac{\beta}{2}\left[\nabla_a\vartheta\nabla^a\vartheta+2V(\vartheta)\right]+\mathcal{L}\MAT\}.
\end{eqnarray}
Here, $g$ stands for the determinant of the metric $g_{ab}$. $R$, $R_{ab}$, $R_{abcd}$, and $\text{}^{*}R_{abcd}$ are the Ricci scalar, Ricci tensor, and the Riemann tensor and its dual, respectively, with the latter defined as 
\be
\text{}^{*}R^a_{~bcd}= \frac{1}{2} \varepsilon_{cd}^{~~ef}R^a_{~bef}\,,
\ee
and $\varepsilon^{abcd}$ the Levi-Civita tensor. The quantity $\mathcal{L}_{\MAT}$ is the external matter Lagrangian, $\vartheta$ is a field, $f_{i}(\vartheta)$ are functionals of this field, $(\alpha_i,\beta)$ are coupling constants, and $\kappa=1/(16\pi)$.

We will focus on two specific theories within QG, EdGB gravity and dCS gravity. In EdGB gravity, $(\alpha_1,\alpha_2,\alpha_3,\alpha_4)=(\alpha_\EdGB,-4\alpha_\EdGB,\alpha_\EdGB,0)$ and $(f_1,f_2,f_3,f_4)=(\vartheta,\vartheta,\vartheta,0)$, where $\alpha_\EdGB$ is the EdGB gravity coupling constant and $\vartheta$ is the dilaton\footnote{In EdGB gravity, the functionals of the dilaton field are actually given by $(f_1,f_2,f_3,f_4)=(e^\vartheta,e^\vartheta,e^\vartheta,0)$. For the BH solution we study in this work we assume $\vartheta$ is at the minimum of its potential, $V(\vartheta)$, and then Taylor expand about small perturbations from the minimum, $f_i(\vartheta)=f_i(0)+f'_i(0)\vartheta+\mathcal{O}(\vartheta^2)$ where $f_i(0)$ and $f'_i(0)$ are constants. The $\vartheta$-independent term leads to a theory with a minimally coupled field, i.e.~the field does not interact with the curvature invariants. The Guass-Bonnet invariant, $R_{\text{GB}}=R^2-4R_{ab}R^{ab}+R_{abcd}R^{abcd}$, is a topological invariant, and thus, the $f_i(0)$ term does not modify the field equations. The $f_i(0)$ term is then irrelevant and we neglect it, instead focusing on the $f'_i(0)$ term, which can be modeled by letting $f_i(\vartheta)=c_i\vartheta$. We reabsorb the constant $c_i$ into the coupling parameter $\alpha_\EdGB$, and then, the functionals are given by $(f_1,f_2,f_3,f_4)=(\vartheta,\vartheta,\vartheta,0)$.}. In dCS gravity, $(\alpha_1,\alpha_2,\alpha_3,\alpha_4)=(0,0,0,\alpha_\dCS/4)$ and $(f_1,f_2,f_3,f_4)=(0,0,0,\vartheta)$, where $\alpha_\dCS$ is the dCS gravity coupling parameter and $\vartheta$ is the dCS (axion like) field. The strongest constraint on EdGB gravity comes from low-mass X-Ray binary observations, $\sqrt{|\alpha_\EdGB|}<1.9\times10^5$cm~\cite{Kent-LMXB}. The strongest constraint on dCS gravity comes from Solar System~\cite{alihaimoud} and tabletop experiments~\cite{PhysRevD.86.044037}, $\sqrt{|\alpha_\dCS|}<10^{13}$cm.

EdGB gravity, and particularly dCS gravity, should be understood as \emph{effective field theories}, i.e.~theories that are only valid to leading order in the coupling parameters $\alpha_{\EdGB}$ and $\alpha_{\dCS}$. This is because one can think of them as arising through a low-energy/low-curvature expansion of some more fundamental gravity theory. For example, dCS gravity arises in the low-energy limit of heterotic superstring theory upon four-dimensional compactification~\cite{Polchinski:1998rr,Alexander:2004xd}, in loop quantum gravity when the Barbero-Immirzi parameter is coupled to matter~\cite{PhysRevD.78.064070,Calcagni:2009xz,PhysRevD.80.104007} and effective field theories of inflation~\cite{PhysRevD.77.123541}. Similarly, EdGB also arises in the low-energy limit of heterotic superstring theory~\cite{Moura:2006pz,lrr-2013-9,Kanti:1995vq,Torii:1996yi,Kanti:1997br}, where the scalar field is here the dilaton. In this sense then, both EdGB and dCS gravity should be considered as an approximate theory valid up to a cutoff energy scale, above which one must account for higher-order derivative operators. An approximate, effective-field-theory treatment of these theories is crucial in order to guarantee that instabilities are not non-linearly generated~\cite{PhysRevD.84.084041,Delsate:2014hba}. Indeed, one can show that typical instabilities, like the Ostrogradski one, are not present when one properly treats the theory as effective~\cite{Burgess:2003jk,lrr-2013-9}.

Let us now consider BH solutions in these theories. In GR, the solution for an isolated, stationary, axisymmetric, and uncharged BH is the Kerr metric. The line element associated with this metric in Boyer-Lindquist coordinates ($t,r,\theta,\phi$) is given by
\begin{eqnarray}
ds_{\K}^2=&-\left(1-\frac{2Mr}{\Sigma_{\K}}\right)dt^2-\frac{4Mar\sin^2\theta}{\Sigma_{\K}}dtd\phi+\frac{\Sigma_{\K}}{\Delta_{\K}}dr^2
\nonumber \\
&+\Sigma_{\K} 
d\theta^2+\left(r^2+a^2+\frac{2Ma^2r\sin^2\theta}{\Sigma_{\K}}\right)\sin^2\theta 
d\phi^2,
\label{eq:Kerr-metric}
\end{eqnarray}
with $\Delta_{\K}\equiv r^2-2Mr+a^2$ and $\Sigma_{\K}\equiv r^2+a^2\cos^2\theta$. Here $M$ is the mass of the BH and $a\equiv J/M$ is the Kerr spin parameter, where $J:=|\vec{J}|$ is the magnitude of the BH spin angular momentum.

For EdGB gravity and dCS gravity we will focus on the approximate, stationary, and axisymmetric solutions that represent slowly-rotating BHs to second order in the spin~\cite{Ayzenberg:2014aka, PhysRevD.86.044037}. These solutions take the form
\begin{equation}
g_{ab}^\QG=g_{ab}^\K+\zeta\left[g^{[0,1]}_{ab} + \chi \; g^{[1,1]}_{ab}+\chi^2 g^{[2,1]}_{ab}\right],
\end{equation}
where $g_{ab}^\K$ is the Kerr metric and $g^{[x,y]}_{ab}$ are the metric deformations due to EdGB gravity or dCS gravity at each order in spin and are given in~\ref{app:QGsols} for completeness. Here $\chi=a/M = \vec{J}/M^{2}$ is the dimensionless spin parameter and $\zeta=16\pi\alpha^2/M^4$ is the dimensionless coupling parameter, where we have set $\beta=1$. 

Due to the slow-rotation and small-coupling ($\zeta<<1$) expansions used to find these BH solutions in EdGB gravity and dCS gravity, they contain spurious features that would not appear in an exact solution. An example of such a spurious feature is that these solutions contain a divergence at $r=2M$, which is unrelated to any physical property of the solutions (see e.g.~\cite{McNees:2015srl}). To eliminate such spurious features one can perform a resummation,~i.e. replace terms in the metric that, if expanded in small rotation, would produce higher order terms in $\chi$, such as $r\rightarrow\Sigma_{\K}^{1/2}$. In principle, there are an infinite number of ways to resum the metric and, since the exact solution of a rotating BH is not known in EdGB gravity and dCS gravity, it is unknown which choice of resummation is the correct one to make. For simplicity, our choice of resummation throughout this work is to treat the slowly-rotating solutions as \textit{exact},~i.e.~to not expand in $\chi$ and $\zeta$ when computing observables, recognizing that the results presented may be different with other choices of resummation when $\chi$ is sufficiently large. An analysis done in GR comparing the Kerr spectrum to a spectrum from a slowly-rotating expansion of Kerr suggests this choice of resummation is accurate up to $\chi\approx0.9$~\cite{Ayzenberg:2016ynm}. 

\section{Properties of the dCS and EdGB BH Solutions}
\label{sec:prop-of-sol}

BH solutions in dCS and EdGB gravity were discussed in detail in~\cite{Ayzenberg:2014aka, PhysRevD.86.044037}. We here discuss the properties of these BHs that are related to continuum spectrum observations, summarizing results from~\cite{Ayzenberg:2014aka, PhysRevD.86.044037} and presenting new results when necessary.

\subsection{Test particle motion}

In order to calculate the continuum spectrum of an accretion disk orbiting a black hole it is first necessary to determine the motion of test particles, i.e.~massive particles with extremely weak self-gravity. In GR, test particles follow geodesics, and the same was proven to be true in dCS gravity~\cite{Sopuerta:2009iy}. We here prove that test particles follow geodesics in EdGB gravity.

We begin from the action of a test particle moving along a worldline $x^a=z^a(\lambda)$, where $\lambda$ parameterizes the trajectory. The action is given by~\cite{lrr-2004-6}
\begin{equation}
S_{\text{mat}}=-m\int_\gamma d\lambda\sqrt{-g_{ab}(z)\frac{dz^a}{d\lambda}\frac{dz^b}{d\lambda}},
\end{equation}
where $m$ is the mass of the test particle and $dz^a/d\lambda$ is the tangent to the worldline $\gamma$. By varying $S_{\text{mat}}$ with respect to the metric we can find the contribution to the matter stress-energy tensor from this test particle. Using that the proper time $\tau$ is related to $\lambda$ via $d\tau=d\lambda\sqrt{-g_{ab}(z)\frac{dz^a}{d\lambda}\frac{dz^b}{d\lambda}}$ and that the particle four-velocity $u^a=dz^a/d\tau$ obeys $g_{ab}u^a u^b=-1$, the matter stress-energy tensor of the test particle can be written as
\begin{equation}
T^{ab}_{\text{mat}}(x^c)=m\int\frac{d\tau}{\sqrt{-\textbf{g}}}u^a u^b\delta^{(4)}[x^{c}-z^{c}(\tau)],
\end{equation}
where $\textbf{g}$ denotes the metric determinant and $\delta^{(4)}$ is the four-dimensional Dirac density defined by $\int d^4x\sqrt{-\textbf{g}} \; \delta^{(4)}(x^{c})=1$. One can easily show that the divergence of $T^{ab}_{\text{mat}}$ is given by
\begin{equation}
\nabla_b T^{ab}_{\text{mat}}=m\int\frac{d\tau}{\sqrt{-\textbf{g}}}\frac{d^2z^a}{d\tau^2}\delta^{(4)}[\textbf{x}-\textbf{z}(\tau)].
\end{equation}

The field equations of EdGB gravity with $V(\vartheta)=0$ and $\beta=1$ are given by~\cite{Ayzenberg:2014aka}
\begin{equation}
G_{ab}+16\pi\alpha_\EdGB\mathcal{D}_{ab}^{(\vartheta)} - 8\pi T_{ab}^{(\vartheta)}=8\pi T_{ab}^{\text{mat}} ,\label{eq:fieldeq}
\end{equation}
where
\begin{equation}
T_{ab}^{(\vartheta)}=\nabla_a\vartheta\nabla_b\vartheta-\frac{1}{2}g_{ab}\nabla_c\vartheta\nabla^c\vartheta,
\end{equation}
is the stress-energy tensor of the scalar field and
\begin{eqnarray}
\mathcal{D}_{ab}^{(\vartheta)}&\equiv-2R\nabla_a\nabla_b\vartheta+2(g_{ab}R-2R_{ab})\nabla^c\nabla_c\vartheta
\nonumber \\
&+8R_{c(a}\nabla^c\nabla_{b)}\vartheta-4g_{ab}R^{cd}\nabla_c\nabla_d\vartheta+4R_{acbd}\nabla^c\nabla^d\vartheta.
\label{D-tensor}
\end{eqnarray}
The scalar field equation is given by
\begin{equation}
\square\vartheta=-\alpha_\EdGB\left(R^2-4R_{ab}R^{ab}+R_{abcd}R^{abcd}\right)=-\alpha R_{\text{GB}}.\label{eq:scalarfieldeq}
\end{equation}

For test particles to follow geodesics, the divergence of $T_{ab}^{\text{mat}}$ must vanish, which means that the divergence of the
second and third terms on the left-hand side of Eq.~\ref{eq:fieldeq} must cancel. Taking the divergence of Eq.~\ref{eq:fieldeq}, we 
then find that test particles follow geodesics if 
\begin{equation}
\nabla_b\mathcal{D}^{ab}=-\frac{1}{2}(\nabla^a\vartheta)R_{\text{GB}},
\label{geod-cond}
\end{equation}
where we have used that the divergence of $T^{ab}_{(\vartheta)}$ is given by
\begin{equation}
\nabla_b T^{ab}_{(\vartheta)}=(\nabla^a\vartheta)(\square\vartheta)=-\alpha_\EdGB(\nabla^a\vartheta)R_{\text{GB}}.
\end{equation}

Let us then evaluate the left-hand side of Eq.~\ref{geod-cond}. Taking the divergence of Eq.~\ref{D-tensor} we have
\begin{eqnarray}
\nabla_b\mathcal{D}^{ab}&=4R\nabla^{[a}\nabla^{b]}\nabla_b\vartheta-8R^{bc}\nabla^{[a}\nabla^{c]}\nabla_b\vartheta
\nonumber \\
&-8R^{ab}\nabla_{[b}\nabla_{c]}\nabla^{c}\vartheta-8\nabla^{[a}R^{c]b}\nabla_b\nabla_c\vartheta
\nonumber \\
&-4R^{abcd}\nabla_d\nabla_c\nabla_b\vartheta-4\nabla_d  R^{abcd}\nabla_c\nabla_b\vartheta.
\end{eqnarray}
Applying the Bianchi identities and the commutation of covariant derivatives gives
\begin{eqnarray}
\nabla_b\mathcal{D}^{ab}&=-2RR^{ab}\nabla_b\vartheta+4R^{ac}R_{bc}\nabla^b\vartheta
\nonumber \\
&+4R_{cd}R^{acbd}\nabla_b\vartheta+4R^{acde}R_{becd}\nabla^b\vartheta.
\end{eqnarray}
Using the definition of the Weyl tensor, $W_{abcd}\equiv R_{abcd}-(g_{a[c}R_{d]b}-g_{b[c}R_{d]a})+1/3Rg_{a[c}g_{d]b}$, we replace the Riemann tensor to get
\begin{eqnarray}
\nabla_b\mathcal{D}^{ab}&=-\frac{1}{3}R^2\nabla^a\vartheta+R_{bc}R^{bc}\nabla^a\vartheta
\nonumber \\
&-4W^{acde}W_{bcde}\nabla^b\vartheta+4W^{acde}W_{bdce}\nabla^b\vartheta.
\end{eqnarray}
Applying the identities $W^{acde}W_{bcde}=1/4g^a_bW^{cdef}W_{cdef}$ and $W^{acde}W_{bdce}=1/8g^a_bW^{cdef}W_{cdef}$ gives
\begin{equation}
\nabla_b\mathcal{D}^{ab}=(\nabla^a\vartheta)(-\frac{1}{3}R^2+R_{bc}R^{bc}-\frac{1}{2}W_{bcde}W^{bcde}).
\end{equation}
Finally, using the definition of the Weyl tensor we find that the divergence of the stress-energy tensor of the scalar field is given by
\begin{eqnarray}
\nabla_b\mathcal{D}^{ab}&=-\frac{1}{2}(\nabla^a\vartheta)(R^2-4R_{bc}R^{bc}+R_{abcd}R^{abcd})
=-\frac{1}{2}(\nabla^a\vartheta)R_{\text{GB}}. \qquad
\end{eqnarray}
Thus, Eq.~\ref{geod-cond} is satisfied, proving that test particles must follow geodesics in EdGB gravity
\begin{equation}
\frac{d^2z^a}{d\tau^2}=0.
\end{equation}
 
\subsection{Conserved Quantities}

All three BH solutions considered within this work are stationary and axisymmetric, and thus, each possesses a timelike and an azimuthal Killing vector, which in turn implies the existence of two conserved quantities: the specific energy $E$ and the $z$-component of the specific angular momentum $L_z$. In the dCS and the EdGB cases, these Killing vectors are approximate, i.e.~they solve the Killing equation to ${\cal{O}}(\zeta,\chi^{2})$, so $E$ and $L_{z}$ are also approximately conserved. 

The definitions of $E$ and $L_z$ and the normalization condition for the $4$-velocity $u^a u_a=-1$ allow us to find the equations of motion for test particles. As discussed above, such particles follow geodesics of the metric in dCS gravity and EdGB gravity. From the definition of $E$ and $L_z$ we find
\begin{eqnarray}
\dot t=&\frac{Eg_{\phi\phi}+L_z g_{t\phi}}{g_{t\phi}^2-g_{tt}g_{\phi\phi}},\label{eqn:tdot}
\qquad
\dot\phi=-\frac{Eg_{t\phi}+L_z g_{tt}}{g_{t\phi}^2-g_{tt}g_{\phi\phi}},
\end{eqnarray}
where the overhead dot represents  a derivative with respect to the affine parameter (proper time for a massive particle). Substituting Eq.~\ref{eqn:tdot} into the normalization condition, we find
\begin{equation}
g_{rr}\dot r^2+g_{\theta\theta}\dot\theta^2=V_{\text{eff}}(r,\theta;E,L_z),
\end{equation}
where we parameterize the four velocity via $u^a=(\dot t,\dot r,\dot\theta,\dot\phi)$ with overhead dots representing derivatives with respect to proper time, and where the effective potential is
\begin{equation}
V_{\text{eff}}\equiv\frac{E^2g_{\phi\phi}+2EL_z g_{t\phi}+L_z^2g_{tt}}{g_{t\phi}^2-g_{tt}g_{\phi\phi}}-1.\label{eqn:Veff}
\end{equation}

Restricting attention to equatorial and circular orbits, we can obtain explicit expressions for the energy and angular momentum as a function of the metric components. Using the stability and circularity conditions $V_{\text{eff}}=0$ and $\partial V_{\text{eff}}/\partial r=0$, and solving for $E$ and $L_z$, we find
\begin{eqnarray}
E&=-\frac{g_{tt}+g_{t\phi}\Omega}{\sqrt{-(g_{tt}+2g_{t\phi}\Omega+g_{\phi\phi}\Omega^2)}},\label{eqn:E}
\\
L_z&=\frac{g_{t\phi}+g_{\phi\phi}\Omega}{\sqrt{-(g_{tt}+2g_{t\phi}\Omega+g_{\phi\phi}\Omega^2)}},
\end{eqnarray}
where the angular velocity of equatorial circular geodesics is defined via
\begin{equation}
\Omega := \frac{d\phi}{dt}=\frac{-g_{t\phi,r}+\sqrt{(g_{t\phi,r})^2-g_{tt,r}g_{\phi\phi,r}}}{g_{\phi\phi,r}}.
\end{equation}
%

\subsection{ISCO}

The ISCO is the stable circular orbit that is closest to the BH event horizon. Any circular orbit inside the ISCO will thus be unstable and any test particle that finds itself there is expected to rapidly plunge and cross the event horizon. Because of this many accretion disk models assume the inner radius of the disk is exactly at the ISCO. This assumption is motivated by physical arguments, simulations, and observational evidence~\cite{2008ApJ...675.1048R, 2008ApJ...687L..25S, 2010MNRAS.408..752P, 2009ApJ...701.1076G, 2010ApJ...718L.117S} and it could, in principle be relaxed. We, however, will retain this assumption throughout this work and leave its relaxation to future studies. 

Since the ISCO is a geometric property of BHs that plays a key role in the continuum spectrum of accretion disks, let us now calculate its location. The ISCO radius can be found by substituting Eq.~\ref{eqn:E} into Eq.~\ref{eqn:Veff}, and then solving $\partial^2 V_{\text{eff}}/\partial r^2=0$ for $r$. The ISCO radius for equatiorial geodesics in Kerr is
\begin{equation}
r_\ISCO=M\left\{3+Z_2\mp\left[\left(3-Z_1\right)\left(3+Z_1+2Z_2\right)\right]^{1/2}\right\},
\end{equation}
where
\begin{eqnarray}
Z_1=&1+\left(1-\chi^2\right)^{1/3}\left[\left(1+\chi\right)^{1/3}+\left(1-\chi\right)^{1/3}\right],
\\
Z_2=&\left(3\chi^2+Z_1^2\right)^{1/2},
\end{eqnarray}
where the $\mp$ denotes whether the disk's angular momentum is in the same $\left(-\right)$ or the opposite $\left(+\right)$ direction as the BH's angular momentum.

\begin{figure*}[hpt]
\includegraphics[width=1.0\columnwidth{},clip=true]{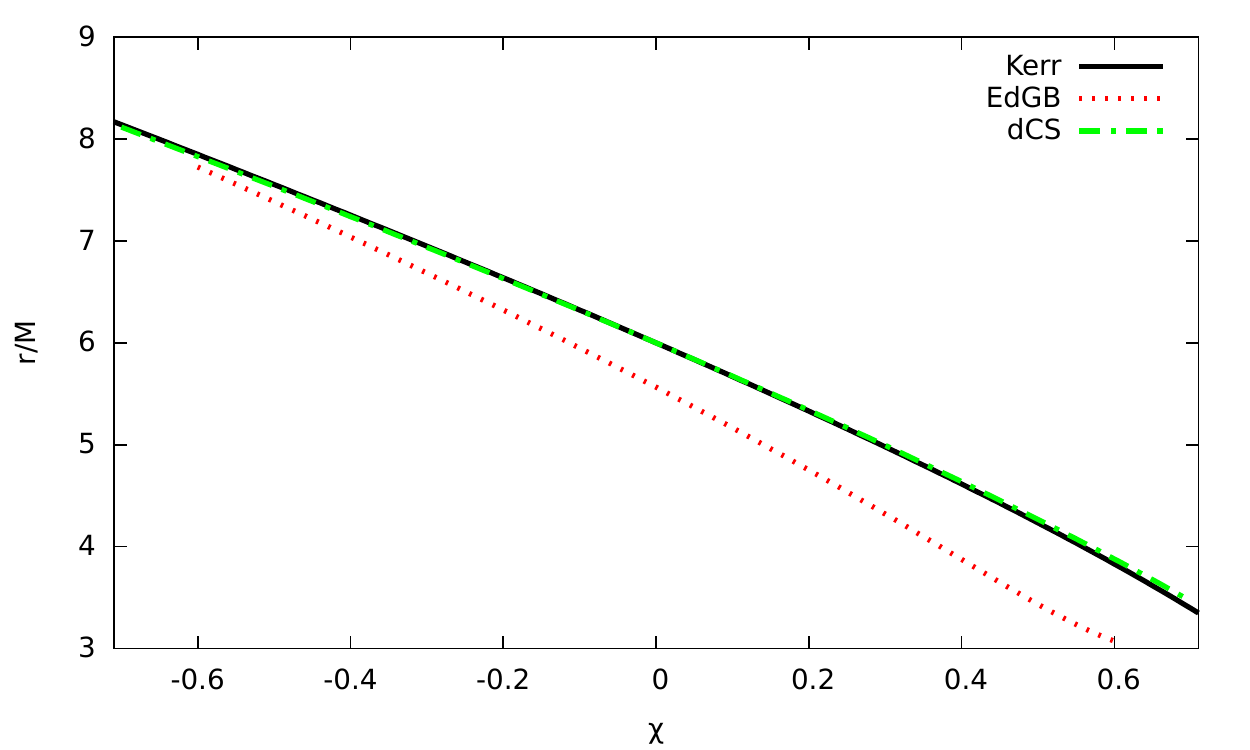}
\caption{(color online) ISCO radius as a function of dimensionless spin $\chi$ for Kerr (black solid line), EdGB gravity with $\sqrt{|\alpha_\EdGB|}=1.9\times10^5$cm (red dotted line), and dCS gravity with $\sqrt{|\alpha_\dCS|}=2.33\times10^5$cm (green dashed-dotted line), for a BH mass $M=5M_\odot$. \label{fig:isco}}
\end{figure*}

The ISCO radius in the EdGB and dCS BH solutions when treated as exact must be solved for numerically as the solutions to $\partial^2 V_{\text{eff}}/\partial r^2=0$ are not analytically tractable. Note that, when the solutions are treated as approximate, the ISCO radius can be found analytically by expanding in $\chi$ and $\zeta$; the difference, however, is negligible in dCS gravity and at most $\sim2\%$ in EdGB gravity for the ranges of $\chi$ and $\zeta$ that we consider. To remain consistent throughout this work we also compute the ISCO radius in Kerr numerically instead of using the above analytic solution, ensuring that the numerical ISCO agrees with the analytic to within our numerical error. Figure~\ref{fig:isco} shows the ISCO radius as a function of dimensionless spin $\chi$ for Kerr, the EdGB gravity solution with $\sqrt{|\alpha_\EdGB|}=1.9\times10^5$cm, and the dCS gravity solution with $\sqrt{|\alpha_\dCS|}=2.33\times10^5$cm, for a BH with mass $M=5M_\odot$\footnote{A choice of larger EdGB coupling would have led to a larger deviation from GR in the ISCO radius, but this is already ruled out by observations. We here focus on how well continuum spectrum observations can place constraints relative to current constraints~\cite{Kent-LMXB}, so our choice of $\alpha_\EdGB$ saturates the latter. Similarly, a choice of smaller dCS coupling would have led to a smaller deviation from GR, with the deviation vanishing for non-spinning BHs. Our choice of $\alpha_\dCS$ maximizes the deviation without violating the small-coupling approximation used to find the dCS BH solution. The current constraint on $\alpha_\dCS$ violates the small-coupling approximation for all BH masses studied in this work.}. Observe that the ISCO location in EdGB BHs is significantly different from the ISCO in Kerr BHs, even in the non-spinning limit, while the deviation in dCS BHs is essentially negligible even at higher spins. 

The amount of deviation in the ISCO radius between the Kerr solution and the solutions in EdGB gravity and dCS gravity is not an intrinsic property of the BH solutions. Non-spinning BHs in EdGB gravity are not given by the Schwarzschild metric, in contrast to non-spinning BHs in dCS gravity. Thus, the ISCO radius in the EdGB BH solution is different from GR even when $\chi=0$, while the ISCO radius in the dCS BH case is the same as in GR. This also means that any modifications to the ISCO radius in rotating dCS BHs must arise from spin-dependent terms in the metric, which are subdominant relative to terms independent of spin for slowly-rotating BHs. The small modification to the ISCO radius in dCS gravity is also due to our choice of coupling constant. A larger coupling would increase the deviation from Kerr, but may violate the small-coupling approximation used in finding the solution studied here.

A measurement of the ISCO radius is thus generically not able to disentangle the effects of spin and the effect of a modification to GR of the type considered here.  That is, given a measurement of ISCO radius, one can always find a value of the spin that will match the ISCO radius for any value of $\alpha_{\dCS}$. Since the radiation that originates mostly near the innermost radius of the disk, assumed here to be the ISCO, dominates the continuum spectrum, this in turn leads to a degeneracy between spin and coupling constant in the continuum spectrum observation. Without a second, independent measurement of the spin it is not possible to test GR and constrain modified theories of gravity with continuum spectrum observations alone.

%

\subsection{Gravitational Redshift}

The gravitational redshift is a quantity that describes how much the frequency of photons changes as they travel out of the BH potential. We can define this quantity via
\begin{equation}
g\equiv\frac{E_o}{E_e}=\frac{(p_a u^a)_o}{(p_b u^b)_e},\label{eqn:redshift1}
\end{equation}
where $p_a$ is the four-momentum of a photon traveling from the emitting material to the observer, and $u^a_o$ and $u^a_e$ are the four-velocities of the observer and the emitting material, respectively. 

Let us now detail how to compute this redshift quantity explicitly in terms of components of the metric tensor by first  focusing on the photon's four-momentum. The BH solutions we study are stationary and axisymmetric, so they are independent of the $t$ and $\phi$ coordinates. Thus, the corresponding components of the four-momentum are conserved and we can write
\begin{equation}
p_a=(p_t,p_r,p_\theta,p_\phi)=(-E,p_r,p_\theta,L_z).
\end{equation}

Let us now focus on the four-velocities of the observer and the emitting material. If we treat the observer as static, then $u_o^a=(1,0,0,0)$. The four-velocity for material in a circular orbit on the equatorial plane is simply 
\begin{equation}
u_e^a=u_e^t(1,0,0,\Omega),
\end{equation}
where
\begin{equation}
u_e^t=\frac{1}{\sqrt{-(g_{tt}+2g_{t\phi}\Omega+g_{\phi\phi}\Omega^2)}}
\end{equation}
to enforce the timelike normalization condition. 

With these quantities computed, we can now explicitly solve for the redshift factor. The numerator of Eq.~\ref{eqn:redshift1} is simply $(p_au^a)_o=-E$, while the denominator is $(p_au^a)_e=-Eu_e^t+u_e^t\Omega L_z$, yielding the redshift factor
\begin{equation}
g=\frac{\sqrt{-(g_{tt}+2g_{t\phi}\Omega+g_{\phi\phi}\Omega^2)}}{1-\Omega\xi},
\end{equation}
where $\xi=L_z/E$.
 
The redshift factor depends on the ratio of the angular momentum and the energy, which although conserved are not directly measurable. We can recast the redshift factor, however, in terms of celestial coordinates $\alpha$ and $\beta$ as follows. First, we place the observer at spatial infinity $(r=+\infty)$ at an inclination angle $\iota$ between the observer's line of sight and the BH's angular momentum. We then define $(\alpha,\beta)$ as the Cartesian coordinates on the observer's plane of the sky, as measured from the observer's line of sight, i.e.~measured in directions perpendicular and parallel to the rotation axis of the BH when projected onto the observer's plane of the sky, respectively. At large spatial distances, and using the fact that the BH metrics are asymptotically flat, the celestial coordinate $\alpha$ is given by
\begin{equation}
\alpha=\lim_{r\rightarrow\infty}-\frac{rp^\phi}{p^t}=\frac{-\xi}{\sin\iota}.
\end{equation}
Neglecting light-bending\footnote{There is no reason to assume the effect of light-bending is negligible compared to other relativistic effects, but incorporating its effects would require a general relativistic ray-tracing code and this is beyond the scope of this work.}, $\alpha$ can also be written as $\alpha=r\cos\phi$ where $\phi=0$ is along the line of nodes, where the disk intersects the observer's plane of the sky. Then, solving for $\xi$ one finds $\xi=-r\cos\phi\sin\iota$, and thus the redshift factor can be written as
\begin{equation}
g=\frac{\sqrt{-(g_{tt}+2g_{t\phi}\Omega+g_{\phi\phi}\Omega^2)}}{1+r\Omega\cos\phi\sin\iota}.\label{eqn:redshift2}
\end{equation}
This expression depends only on the metric components, the angle $\phi$ and the material's angular velocity. 

\section{Continuum Spectrum in dCS and EdGB BHs}
\label{sec:spec}

We use the Novikov-Thorne accretion disk model~\cite{1973blho.conf..343N}, the standard general relativistic model for geometrically-thin and optically-thin accretion disks. The model assumes the disk is in the equatorial plane of the BH and the disk particles move on nearly geodesic circular orbits, i.e.~geodesics except for a small radial momentum. With these assumptions, two of the equations that describe the time-averaged radial structure of the disk and are that are used for calculating the continuum spectrum are
\begin{eqnarray}
\dot M&=-2\pi\sqrt{-\textbf{g}}\Sigma(r)u^r=\text{constant},
\\
\mathcal{F}(r)&=\frac{\dot M}{4\pi\sqrt{-\textbf{g}}}f(r),\label{eqn:flux}
\end{eqnarray}
where $\dot M$ and $\mathcal{F}$ are the time-averaged mass accretion rate and radially-dependent energy flux, respectively. In these equations, $\Sigma(r)$ is the surface density, $u^r$ is the radial four-velocity of the disk particles, $\textbf{g}$ is the determinant of the metric in the near equatorial plane in cylindrical coordinates, and the function $f(r)$ is defined by
\begin{equation}
f(r)=\frac{-\partial_r\Omega}{(E-\Omega L_z)^2}\int^{r}_{r_\text{in}}(E-\Omega L_z)(\partial_{r'}L_z)dr'.
\end{equation}
Here $r_\text{in}$ is the inner radius of the accretion disk, which we choose to be the location of the ISCO.

The accretion rate $\dot M$ can be rewritten as $\dot M=L_{\text{bol}}/\eta$, where $L_{\text{bol}}$ is the bolometric luminosity and $\eta$ is the radiative efficiency, the efficiency of conversion between rest-mass and EM energy. The radiative efficiency can be written as $\eta=1-E(r_\ISCO)$, by assuming the energy radiated by a particle falling into a BH is approximately equal to the binding energy of the ISCO~\cite{1973grav.book.....M}. The accretion rate is then given by 
\begin{equation}
\dot M=\frac{L_{\text{bol}}}{1-E(r_\ISCO)},
\end{equation}
and the radial energy flux of Eq.~\ref{eqn:flux} can be rewritten as
\begin{eqnarray}
\mathcal{F}(r)=&\frac{L_\text{bol}}{4\pi\sqrt{-\textbf{g}}\left[1-E(r_\ISCO)\right]}\frac{-\partial_r\Omega}{\left(E-\Omega L_z\right)^2}
\nonumber \\
&\times\int^r_{r_\text{in}}\left(E-\Omega L_z\right)\left(\partial_{r'}L_z\right)dr'.
\label{eqn:Eflux}
\end{eqnarray}

Assuming the disk is in thermal equilibrium and modeling the radiation emitted by the disk as a black-body, we can compute its luminosity. This quantity is nothing but the integral of the spectral radiance given by Planck's law over the extent of the disk, namely
\begin{equation}
L(\nu)=\frac{8\pi h}{c^2}\cos\iota\int^{r_{\text{out}}}_{r_{\text{in}}}\int^{2\pi}_{0}\frac{\nu^3\sqrt{-\bf{g}}}{\exp{[h\nu/gk_{\text{B}}T(r)]}-1}drd\phi,\label{eqn:lum}
\end{equation}
where $g$ is the redshift found in Eq.~\ref{eqn:redshift2}, $h$ is the Planck constant, $k_{\text{B}}$ is the Boltzmann constant, $\nu$ is the observed frequency, $r_{\text{out}}$ is the outer radius of the disk, and we have here restored the speed of light $c$. The quantity $T(r)$ is the temperature of the disk, which can be related to the radial energy flux using the Stefan-Boltzmann law
\begin{equation}
T(r)=\left(\frac{\mathcal{F}(r)}{\sigma}\right)^{1/4},
\label{eqn:SB}
\end{equation}
where $\sigma$ is the Stefan-Boltzmann constant.

The main observable we will be concerned with is the accretion disk luminosity, which as we can see depends on the metric in various ways. The luminosity $L(\nu)$ is given by Eq.~\ref{eqn:lum}, which depends on the metric via the  determinant factor in its integrand, the ISCO in the limits of integration, and also through the temperature $T(r)$. The latter is given in terms of the radial energy flux in Eq.~\ref{eqn:SB}, while the energy flux is given in Eq.~\ref{eqn:Eflux}. This flux clearly depends on the metric through its associated conserved quantities $E$ and $L_z$, as well as the angular velocity $\Omega$ of test particles in a circular orbit. It stands to reason, then, that if the metric changes, for example if modified gravity theories do not allow the Kerr metric as a solution for BH spacetimes, then the luminosity of its associated accretion disk will also change. 

\subsection{Method}

We wish to determine whether better-than-current constraints can be placed on EdGB gravity and dCS gravity using current and next generation continuum spectrum observations. Following the same prescription as~\cite{Ayzenberg:2016ynm}, let us assume that the Kerr metric is the correct description of a BH spacetime and that the associated spectrum is our observation. We shall refer to the Kerr spectrum observation as the \textit{injected synthetic signal} or \textit{injection} for short. Further, we use the spectrum calculated with the EdGB gravity or dCS gravity metrics as our \textit{model} and fit it to the injection. 

When constructing both the injections and the models, the spectrum is calculated using Eq.~\ref{eqn:lum}, as explained in the previous subsection. The integrals are numerically evaluated using Simpson's rule with step sizes chosen to ensure numerical error is small. For the energy flux integration, we choose a radial step size of $\delta r=0.1M$, with a much smaller step size of $\delta r=10^{-4}M$ for $r\leq r_\ISCO+2.5M$,  as the energy flux changes rapidly near the ISCO radius. For the luminosity integration, we choose step sizes of $\delta r=M$ and $\delta\phi=0.1$. A lengthy numerical investigation was performed to guarantee the numerical error is under control with these choices of step sizes.

The parameters of the spectrum model outlined in Sec.~\ref{sec:spec} are $\vec{\lambda}=\left(M,\chi,\iota,F_{\text{bol}}\right)$, i.e.~the BH mass, its dimensionless spin $\chi = a/M$, the inclination angle, and the bolometric luminosity, respectively. The latter, $F_{\text{bol}}$, should in principle be extracted from observations, but since we wish to focus on the impact of modified BH solutions rather than the properties of the accretion disk itself, we will fix $L_\text{bol}=1.2572\times10^{36}$ erg/s. This luminosity is also equal to $10\%$ of the Eddington luminosity, $L_{\text{Edd}}=1.2572\times10^{38}\left(M/M_{\odot}\right)$erg/s for a $1 M_\odot$ object. This leaves the mass $m$, the spin $\chi$, and the inclination angle $\iota$ as the parameters of the spectrum model, for all of which we choose a flat prior over the following ranges. For the mass and inclination angle, we choose ranges that are representative of current BH continuum spectrum observations: $6M_\odot\leq M\leq19M_\odot$ and $10^\circ\leq\iota\leq80^\circ$. The spin range is limited by the region of validity for the slow-rotating EdGB gravity and dCS gravity solutions. For EdGB gravity we use the range $-0.6\leq\chi\leq0.6$~\cite{Ayzenberg:2014aka} and for dCS gravity we use the range $-0.7\leq\chi\leq0.7$~\cite{Yagi:2012vf}.

As we wish to compare the projected constraints we will obtain against current constraints on EdGB gravity and dCS gravity we fix the coupling constant $\alpha$ in each model, thus not including it as a parameter of the model. In the case of EdGB gravity we choose $\sqrt{|\alpha_\EdGB|}=1.9\times10^5$cm, which saturates the current constraint~\cite{Kent-LMXB}. For dCS gravity we choose $\sqrt{|\alpha_\dCS|}=2.33\times10^5$cm, which gives a dimensionless coupling of $\zeta=0.5$ for BH mass $M=5M_\odot$, the smallest mass, and thus, the largest dimensionless coupling, used in our analysis. For coupling values larger than this, the small-coupling approximation used to construct the dCS BH solution would be violated. In both dCS gravity and EdGB gravity, using a smaller coupling parameter would lead to a smaller deviation from GR, but as we are comparing against current constraints we choose values that maximize the deviation.

We estimate parameters in the model by minimizing the relative $\chi^2$ over all parameters. The reduced $\chi^2$ is defined as
\begin{equation}
\chi^2_{\text{red}}=\frac{\chi^2}{N}=\frac{1}{N}\sum^N_{i=1}\left[\frac{L_{\text{QG}}(\nu_i,M,\chi,\iota)-L_{\text{K}}(\nu_i,M^*,\chi^*,\iota^*)}{\sigma(\nu_i)}\right]^2,
\end{equation}
where the summation is over $N$ sampling frequencies $\nu_i\in(10^{15},10^{18})$Hz wth 10 samples per decade spaced logarithmically. This sampling choice is representative of that made in the observed spectra of BHs with estimated spins~\cite{2009ApJ...701.1076G, 2008ApJ...679L..37L, 2014ApJ...793L..29S, 2009ApJ...701.1076G, 2010ApJ...718L.122G, 2014ApJ...784L..18M, 2006ApJ...636L.113S, 2011MNRAS.416..941S}. The quantity $L_{\text{QG}}(\nu,M,\chi,\iota)$ is the spectrum model, which depends on the frequency $\nu$ and the model parameters $(M,\chi,\iota)$, while $L_{\text{K}}(\nu,M^*,\chi^*,\iota^*)$ is the injection, which depends on the frequency and the injected parameters $(M^*,\chi^*$, $\iota^*)$. The values of model parameters that minimize the reduced $\chi^2$ are the best-fit model parameters.

We model the standard deviation of the distribution, $\sigma$, via
\begin{equation}
\sigma(\nu_i)=\sigma_M(\nu_i)+\sigma_\chi(\nu_i)+\sigma_\iota(\nu_i),
\end{equation}
where
\begin{eqnarray}
\sigma_M(\nu_i)=&\frac{|L_{\text{K}}(\nu_i,M^*+\delta m,\chi^*,\iota^*)-L_{\text{K}}(\nu_i,M^*-\delta m,\chi^*,\iota^*)|}{2},
\\
\sigma_\chi(\nu_i)=&\frac{|L_{\text{K}}(\nu_i,M^*,\chi^*+\delta\chi,\iota^*)-L_{\text{K}}(\nu_i,M^*,\chi^*-\delta\chi,\iota^*)|}{2},
\\
\sigma_\iota(\nu_i)=&\frac{|L_{\text{K}}(\nu_i,M^*,\chi^*,\iota^*+\delta\iota)-L_{\text{K}}(\nu_i,M^*,\chi^*,\iota^*-\delta\iota)|}{2},
\end{eqnarray}
where $M^*,\chi^*$, and $\iota^*$ are the injected mass, spin, and inclination angle of the Kerr spectrum, respectively. The quantities $\delta M, \delta\chi$, and $\delta\iota$ serve as a way to represent the observational error in the observations. When considering the ability of current telescopes to place constraints on modified gravity, we choose $(\delta M,\delta\chi,\delta\iota)=(1 M_\odot,0.1,1^\circ)$, which is comparable to or better than the error in current BH mass, spin, and inclination angle measurements for BHs in which the spins were measured using continuum spectrum observations~\cite{2015PhR...548....1M}. When considering the ability of next generation telescopes to place constraints on modified theories, we reduce the observational error in the spin parameter by an order of magnitude, i.e.~$\delta\chi=0.01$~\cite{Takahashi:2014cea, Miller:2014cda, Zhang:2016ach}.

\subsection{Results}

We first wish to determine if better-than-current constraints can be placed on modified gravity theories with continuum spectrum observations using current telescopes. To do so we define the weighted deviation $\Delta_A=|A^*-A|/\sigma_A$ where $A=[M,\chi,\iota]$, i.e.~the difference between the value of the injected parameter and the value of the best fit parameter weighted by the error in that parameter. When $\Delta_A>1$, we expect the deviation in the continuum spectrum due to the modified gravity solution to be in principle detectable, i.e.~larger than the observational error, and the modified theory may be constrained. However, if $\Delta_A<1$ the deviation in the continuum spectrum is not detectable (not even in principle) and the modified theory cannot be constrained. 

\begin{figure*}[hpt]
\includegraphics[width=1.0\columnwidth{},clip=true]{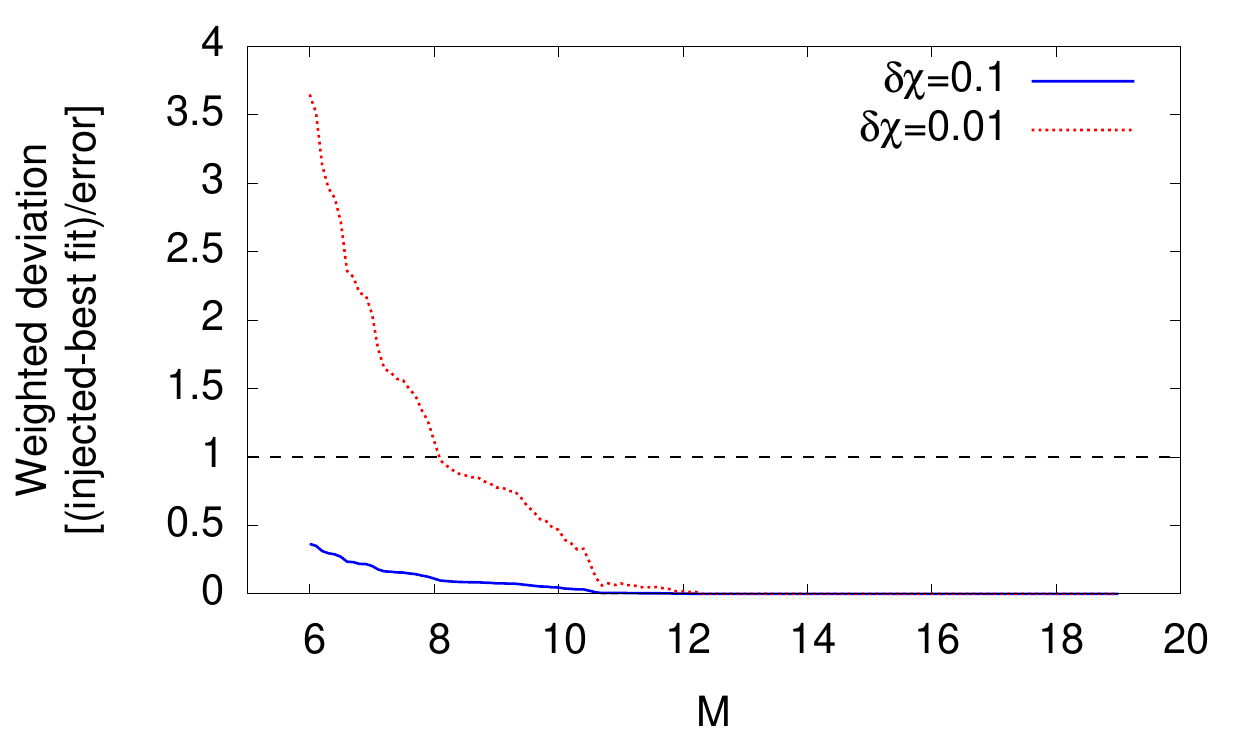}
\caption{(color online) Weighted deviation for spin as a function of injected mass when averaged over injected spin and inclination angle for EdGB gravity with $\sqrt{|\alpha_\EdGB|}=1.9\times10^5$cm, $\delta M=1M_\odot$, and $\delta\iota=1^\circ$. To represent current telescopes we use $\delta\chi=0.1$ (blue solid line) and to represent next generation telescopes we use $\delta\chi=0.01$ (red dotted line). The black dashed line at a weighted deviation of 1 marks the boundary between a deviation being detectable $(> 1)$ and not detectable $(< 1)$, i.e. being able to place a constraint or not place a constraint on a modified theory, respectively. \label{fig:EdGB gravity}}
\end{figure*}

Figure~\ref{fig:EdGB gravity} shows the weighted deviation for spin as a function of injected mass when averaged over the injected spin and inclination angle for EdGB gravity with current telescopes. The weighted deviation for mass and inclination angle, as well as that for mass as a function of injected spin and inclination angle, are approximately zero in the entire range explored, so we do not show them here. Although the weighted deviation is below unity for all parameters, and thus, the deviation due to EdGB gravity is not detectable, the spin weighted deviation does increase as the BH mass decreases. This occurs because the deviation from GR is proportional to the dimensionless coupling $\zeta$, which goes as $1/M^4$, thus deviations are larger in smaller mass BHs. 

The weighted deviations for dCS gravity is also approximately zero, but for the full range of injected masses, spins, and inclination angles we considered. This means the deviation in the metric due to dCS gravity is not detectable with current telescopes at all. The deviation is so small in this case because Schwarzschild is already a solution in dCS gravity, and thus, the non-spinning part of the BH solution is not modified. 

Let us now consider constraints one can place on modified gravity with next generation X-Ray telescopes. The results for dCS gravity are similar to those for current telescopes; the weighted deviations for mass, spin, and inclination angle are all approximately zero for the full range of injected masses, spins, and inclination angles considered. Thus, even with next generation telescopes, dCS gravity cannot be better constrained using these observations. In the EdGB case, however, the situation is slightly different. Figure~\ref{fig:EdGB gravity} shows the weighted deviation for spin as a function of injected mass for EdGB gravity; all other weighted deviations remain below unity, and we thus do not show them here. As the weighted deviation for spin is significantly above 1 for low BH mass, next generation telescopes may be able to place better-than-current constraints on the EdGB gravity coupling constant with continuum spectrum observations of BHs provided $M\lesssim8M_\odot$. As explained previously, deviations from GR are larger for smaller mass BHs because the deviations are proportional to $1/M^4$.

%

\section{Discussion}
\label{sec:disc}

We have studied whether it is possible to place better-than-current constraints on coupling constants in modified gravity theories using BH continuum spectrum observations with both current and next generation X-Ray telescopes. We focused on EdGB gravity and dCS gravity, two theories within the broader class of quadratic gravity theories, as examples of well-motivated modified theories. The BHs were modeled using approximate solutions in EdGB gravity and dCS gravity that are quadratic in the angular momentum and linear in the coupling. We have found that dCS gravity cannot be constrained using continuum spectrum observations, both with current and next generation telescopes, as the modifications in the spectrum are much smaller than the sensitivity of the telescopes. In the EdGB gravity case, however, we find that although current telescopes cannot place better-than-current constraints on the coupling, next generation telescopes will be able to do so provided the BHs observed have a sufficiently small mass.

While our results show that the deviation due to EdGB gravity in the spin parameter extracted from the continuum spectrum is larger than the sensitivity of next generation telescopes, to actually place constraints on $\alpha_\EdGB$ a second, independent measurement of the spin parameter is required to break the degeneracy between the spin and the modified theory. Possible independent measurements include the K$\alpha$ iron line emitted from accretion disks~\cite{Bambi:2016sac} or quasi-periodic oscillations observed in accretion disks~\cite{Motta:2016vwf}. An interesting extension of our work would be to determine if a constraint could still be placed on EdGB gravity with next generation telescopes when a second independent measurement, and the error associated with it, is taken into account.

Our analysis assumes the accretion disk and continuum spectrum are well understood and well modeled, i.e.~the systematic error due to our lack of understanding of accretion disk physics is negligible. In reality this is currently not the case as there are numerous accretion disk models, analytic and numerical, that include different assumptions about the initial conditions and physics involved, and current observations of accretion disks are not able to distinguish between all the models~\cite{Davis:2006cm, lrr-2013-1, Maccarone:2013tea}. The systematic error due to accretion disk model uncertainty is generally estimated to be on the order of the current observational error in continuum spectrum observations, i.e.~$\sim10\%$ error in the accuracy to which BH properties can be estimated~\cite{2015PhR...548....1M}. Before constraints can be placed on modified gravity theories using continuum spectrum observations with next generation telescopes, as our analysis suggests is possible, the systematic error in the model must be brought down to similar levels as the observational error, i.e.~$\sim1\%$ error in the recovered BH parameters.

The approximate BH solutions in EdGB gravity and dCS gravity we studied in this work are of quadratic order in the spin angular momentum and linear order in the dimensionless coupling constant. Repeating the analysis done here with solutions that include higher orders in spin would allow this work to be extended to BHs with higher spins, eventually approaching the maximal spin limit. An extension of this sort that includes higher order in spin effects is particularly important for dCS gravity as the modifications to GR from dCS gravity are only present for rotating BHs. For EdGB gravity a solution that is fifth order in the spin and seventh order in the coupling parameter $\alpha_\EdGB$ was found in~\cite{Maselli:2015tta}, and can be used to extend the work in this paper. For dCS gravity a higher order solution has not yet been found.

Other extensions of our work could be to relax some of the other assumptions that were made, in turn further verifying our conclusions. The effect of light-bending was neglected and could be taken into account with a ray-tracing algorithm. The assumption that the inner radius of the accretion disk is at the ISCO is an important one, but it is not necessarily correct. The Novikov-Thorne model is a simple accretion disk model and real accretion disks are almost certainly more complex. Other accretion disk models with varying inner radii could be used to determine whether our results are independent of accretion disk model.

\ack

We thank Kent Yagi, Alejandro C\'ardenas-Avenda\~no, and Andrew Sullivan for useful discussions. This work was supported by the NSF CAREER Grant PHY-1250636. Some calculations used the computer algebra-systems MAPLE, in combination with the GRTENSORII package~\cite{grtensor}. 

\appendix

\section{BH Solutions in EdGB gravity and dCS gravity}
\label{app:QGsols}

We here provide the metric modifications to the Kerr solution due to EdGB gravity and dCS gravity.

\setlength{\mathindent}{0pt}

In EdGB gravity the only nonvanishing terms are
\begin{eqnarray}
g_{tt}^{[0,1]}&=-\frac{M^3}{r^4}\left[1+\frac{26M}{r}+\frac{66}{5}\frac{M^2}{r^2}+\frac{96}{5}\frac{M^3}{r^3}-\frac{80M^4}{r^4}\right],
\\
g_{rr}^{[0,1]}&=-\frac{M^2}{r^2f^2}\left[1+\frac{M}{r}+\frac{52}{3}\frac{M^2}{r^2}+\frac{2M^3}{r^3}+\frac{16}{5}\frac{M^4}{r^4}-\frac{368}{3}\frac{M^5}{r^5}\right],
\\
g_{t\phi}^{[1,1]}&=\frac{3}{5}\frac{M^4\sin^2\theta}{r^3}\left[1+\frac{140}{9}\frac{M}{r}+\frac{10M^2}{r^2}+\frac{16M^3}{r^3}-\frac{400}{9}\frac{M^4}{r^4}\right],
\\
g_{tt}^{[2,1]}&=-\frac{4463}{2625}\frac{M^3}{r^3}\left[\left(1+\frac{M}{r}+\frac{27479}{31241}\frac{M^2}{r^2}-\frac{2275145}{187446}\frac{M^3}{r^3}-\frac{2030855}{93723}\frac{M^4}{r^4}\right.\right.
\nonumber \\
&\left.-\frac{99975}{4463}\frac{M^5}{r^5}+\frac{1128850}{13389}\frac{M^6}{r^6}+\frac{194600}{4463}\frac{M^7}{r^7}-\frac{210000}{4463}\frac{M^8}{r^8}\right)\left(3\cos^2\theta-1\right)
\nonumber \\
&\left.-\frac{875}{8926}\left(1+\frac{14M}{r}+\frac{52}{5}\frac{M^2}{r^2}+\frac{1214}{15}\frac{M^3}{r^3}+\frac{68M^4}{r^4}+\frac{724}{5}\frac{M^5}{r^5}-\frac{11264}{15}\frac{M^6}{r^6}\right.\right.
\nonumber \\
&\left.\left.+\frac{160}{3}\frac{M^7}{r^7}\right)\right],
\\
g_{rr}^{[2,1]}&=-\frac{M^3}{r^3f^3}\left[\frac{4463}{2625}\left(1-\frac{5338}{4463}\frac{M}{r}-\frac{59503}{31241}\frac{M^2}{r^2}-\frac{7433843}{187446}\frac{M^3}{r^3}+\frac{13462040}{93723}\frac{M^4}{r^4}\right.\right.
\nonumber \\
&\left.\left.-\frac{7072405}{31241}\frac{M^5}{r^5}+\frac{9896300}{13389}\frac{M^6}{r^6}-\frac{28857700}{13389}\frac{M^7}{r^7}+\frac{13188000}{4463}\frac{M^8}{r^8}-\frac{7140000}{4463}\frac{M^9}{r^9}\right)\right.
\nonumber \\
&\left.\times\left(3\cos^2\theta-1\right)-\frac{r}{2M}\left(1-\frac{M}{r}+\frac{10M^2}{r^2}-\frac{12M^3}{r^3}+\frac{218}{3}\frac{M^4}{r^4}+\frac{128}{3}\frac{M^5}{r^5}-\frac{724}{15}\frac{M^6}{r^6}\right.\right.
\nonumber \\
&\left.\left.-\frac{22664}{15}\frac{M^7}{r^7}+\frac{25312}{15}\frac{M^8}{r^8}+\frac{1600}{3}\frac{M^9}{r^9}\right)\right],
\\
g_{\theta\theta}^{[2,1]}&=-\frac{4463}{2625}\frac{M^{3}}{r}\left(1+\frac{10370}{4463}\frac{M}{r}+\frac{266911}{62482}\frac{M^2}{r^2}+\frac{63365}{13389}\frac{M^3}{r^3}-\frac{309275}{31241}\frac{M^4}{r^4}\right.
\nonumber \\
&\left.-\frac{81350}{4463}\frac{M^5}{r^5}-\frac{443800}{13389}\frac{M^6}{r^6}+\frac{210000}{4463}\frac{M^7}{r^7}\right)\left(3\cos^2\theta-1\right),
\\
g_{\phi\phi}^{[2,1]}&=g_{\theta\theta}^{[2,1]}\sin^2\theta,
\end{eqnarray}
where $f=1-2M/r$.

In dCS gravity the only nonvanishing terms are
\begin{eqnarray}
g_{t\phi}^{[1,1]}&=\frac{5}{8}\frac{M^5}{r^4}\left(1+\frac{12}{7}\frac{M}{r}+\frac{27}{10}\frac{M^2}{r^2}\right)\sin^2\theta,
\\
g_{tt}^{[2,1]}&=\frac{M^3}{r^3}\left[\frac{201}{1792}\left(1+\frac{M}{r}+\frac{4474}{4221}\frac{M^2}{r^2}-\frac{2060}{469}\frac{M^3}{r^3}+\frac{1500}{469}\frac{M^4}{r^4}-\frac{2140}{201}\frac{M^5}{r^5}\right.\right.
\nonumber \\
&\left.+\frac{9256}{201}\frac{M^6}{r^6}-\frac{5376}{67}\frac{M^7}{r^7}\right)\left(3\cos^2\theta-1\right)
\nonumber \\
&\left.-\frac{5}{384}\frac{M^2}{r^2}\left(1+\frac{100M}{r}+\frac{194M^2}{r^2}+\frac{2220}{7}\frac{M^3}{r^3}-\frac{1512}{5}\frac{M^4}{r^4}\right)\right],
\\
g_{rr}^{[2,1]}&=\frac{M^3}{r^3f^2}\left[\frac{201}{1792}f\left(1+\frac{1459}{603}\frac{M}{r}+\frac{20000}{4221}\frac{M^2}{r^2}+\frac{51580}{1407}\frac{M^3}{r^3}-\frac{7580}{201}\frac{M^4}{r^4}\right.\right.
\nonumber \\
&\left.-\frac{22492}{201}\frac{M^5}{r^5}-\frac{40320}{67}\frac{M^6}{r^6}\right)\left(3\cos^2\theta-1\right)
\nonumber \\
&\left.-\frac{25}{384}\frac{M}{r}\left(1+\frac{3M}{r}+\frac{322}{5}\frac{M^2}{r^2}+\frac{198}{5}\frac{M^3}{r^3}+\frac{6276}{175}\frac{M^4}{r^4}-\frac{17496}{25}\frac{M^5}{r^5}\right)\right],
\\
g_{\theta\theta}^{[2,1]}&=\frac{201}{1792}\frac{M^3}{r}\left(1+\frac{1420}{603}\frac{M}{r}+\frac{18908}{4221}\frac{M^2}{r^2}+\frac{1480}{603}\frac{M^3}{r^3}+\frac{22460}{1407}\frac{M^4}{r^4}\right.
\nonumber \\
&\left.+\frac{3848}{201}\frac{M^5}{r^5}+\frac{5376}{67}\frac{M^6}{r^6}\right)\left(3\cos^2\theta-1\right),
\\
g_{\phi\phi}^{[2,1]}&=\sin^2\theta g_{\theta\theta}^{[2,1]}.
\end{eqnarray}

\section*{References}
\bibliography{biblio}
\bibliographystyle{iopart-num}

\end{document}